\documentclass[runningheads]{llncs}
\usepackage[utf8]{inputenc}
\usepackage{isabelle,isabellesym}
\usepackage{amsmath}
\usepackage{amsfonts}
\usepackage{amssymb}
\usepackage{bbm}  % Para el \bb{1}
\usepackage{multidef}
\usepackage{verbatim}
\usepackage{stmaryrd} %% para \llbracket
\usepackage{hyperref}
\usepackage{xcolor}
\usepackage{framed}

\newcommand{\lsii}{\longleftrightarrow}

\DeclareMathOperator{\dom}{domain}

\DeclareMathOperator{\Con}{Con}

\newcommand{\modelo}[1]{\mathbf{#1}}
\newcommand{\axiomas}[1]{\mathit{#1}}
\newcommand{\clase}[1]{\mathsf{#1}}

\newcommand{\operador}[1]{\mathbf{#1}}

\multidef{\clase{#1}}{Card,HC,HF,Lim,On->Ord,Reg,WF,Ord}

\DeclareMathAlphabet{\mathbbm}{U}{bbm}{m}{n} 

\newcommand{\PP}{\mathbbm{P}}

\multidef[prefix=cal]{\mathcal{#1}}{A-Z}
\multidef{\modelo{#1}}{A,BB->B,CC->C,NN->N,QQ->Q,RR->R,ZZ->Z}

\multidef[prefix=p]{\mathbb{#1}}{A-Z}

\renewcommand{\H}{\operador{H}}
\renewcommand{\S}{\operador{S}}

\renewcommand{\emptyset}{\varnothing}

\newcommand{\Pow}{\mathop{\mathcal{P}}}
\renewcommand{\P}{\Pow}

\newcommand{\lb}{\langle}
\newcommand{\rb}{\rangle}

\renewcommand{\phi}{\varphi}

\newcommand{\defi}{\mathrel{\mathop:}=}
\newcommand{\forces}{\Vdash}

\newcommand{\sig}{\ensuremath{\sigma}}

\newcommand{\om}{\ensuremath{\omega}}
\newcommand{\AC}{\axiomas{AC}}

\newcommand{\CH}{\axiomas{CH}}
\newcommand{\ZFC}{\axiomas{ZFC}}
\newcommand{\ZF}{\axiomas{ZF}}

\newcommand{\bx}{\bar{x}}

\newcommand{\union}{\mathop{\textstyle\bigcup}}

\newcommand{\sbq}{\subseteq}

\DeclareMathOperator{\val}{\mathit{val}}

\renewcommand{\PP}{\mathbb{P}}

\newcommand{\formula}{\isatt{formula}}

\newcommand{\forceisa}{\mathop{\mathtt{forces}}}

\newcommand{\frecR}{\mathrel{\isatt{frecR}}}
\newcommand{\forceseq}{\mathop{\isatt{forces{\isacharunderscore}eq}}}
\newcommand{\forcesmem}{\mathop{\isatt{forces{\isacharunderscore}mem}}}
\newcommand{\forcesat}{\mathop{\isatt{forces{\isacharunderscore}at}}}
\newcommand{\pleq}{\preceq}

\makeatletter
\@ifpackageloaded{txfonts}\@tempswafalse\@tempswatrue
\if@tempswa
  \DeclareFontFamily{U}{txsymbols}{}
  \DeclareFontFamily{U}{txAMSb}{}
  \DeclareSymbolFont{txsymbols}{OMS}{txsy}{m}{n}
  \SetSymbolFont{txsymbols}{bold}{OMS}{txsy}{bx}{n}
  \DeclareFontSubstitution{OMS}{txsy}{m}{n}
  \DeclareSymbolFont{txAMSb}{U}{txsyb}{m}{n}
  \SetSymbolFont{txAMSb}{bold}{U}{txsyb}{bx}{n}
  \DeclareFontSubstitution{U}{txsyb}{m}{n}
  \DeclareMathSymbol{\aleph}{\mathord}{txsymbols}{64}
  \DeclareMathSymbol{\beth}{\mathord}{txAMSb}{105}
  \DeclareMathSymbol{\gimel}{\mathord}{txAMSb}{106}
  \DeclareMathSymbol{\daleth}{\mathord}{txAMSb}{107}
\fi
\makeatother

\newcommand{\quantRel}[3]{#1 #2\kern -1pt[#3]}

\newif\ifarXiv
\newif\ifIEEE

\usepackage{graphicx}
\hypersetup{
  colorlinks,
  urlcolor={blue},
  linkcolor={blue!50!black},
  citecolor={blue!50!black},
}
\begin{document}
\title{Formalization of Forcing in Isabelle/ZF%
  \thanks{Supported by Secyt-UNC project 33620180100465CB and Conicet.}%
}
\author{Emmanuel Gunther\inst{1} \and
Miguel Pagano\inst{1} \and \\
Pedro Sánchez Terraf\inst{1,2}%\orcidID{0000-0003-3928-6942}
}
\authorrunning{E.~Gunther, M.~Pagano, P.~Sánchez Terraf}
%\authorrunning{E. Gunther et al.}
\institute{Universidad Nacional de C\'ordoba. 
  \\  Facultad de Matem\'atica, Astronom\'{\i}a,  F\'{\i}sica y
    Computaci\'on. \\
\email{\{gunther,pagano,sterraf\}@famaf.unc.edu.ar}
\and
    Centro de Investigaci\'on y Estudios de Matem\'atica (CIEM-FaMAF),
    Conicet. C\'ordoba. Argentina.
}
\maketitle              % typeset the header of the contribution
\begin{abstract}
We formalize the theory of forcing in the set theory framework of
Isabelle/ZF. Under the assumption of the existence of a countable
transitive model of $\ZFC$, we construct a proper generic extension and show
that the latter also satisfies $\ZFC$. In doing so, we remodularized
Paulson's \verb|ZF-Constructibility| library.
\keywords{forcing \and Isabelle/ZF \and countable transitive models
  \and absoluteness \and generic extension \and constructibility.}
\end{abstract}
\section{Introduction}
\label{sec:introduction}

The present work reports on the third stage of our project of
formalizing the theory of forcing and its applications as presented in
one of the more important references on the subject, Kunen's Set
Theory \cite{kunen2011set} (a rewrite of the classical book~\cite{kunen1980}).

We
work using the  implementation of Zermelo-Fraenkel ($\ZF$)
set theory \emph{Isabelle/ZF} by Paulson and Grabczewski \cite{DBLP:journals/jar/PaulsonG96}. In
an early paper \cite{2018arXiv180705174G}, we set up the first
elements of the countable transitive model (ctm) approach, defining
forcing notions, names, generic extensions, and
showing the existence of generic filters via the Rasiowa-Sikorski
lemma (RSL). In our second (unpublished) technical report
\cite{2019arXiv190103313G} we advanced by presenting the first accurate
\emph{formal abstract} of the Fundamental Theorems of Forcing, and
using them to show that that the $\ZF$ axioms apart from Replacement
and Infinity hold in all generic extensions.

This paper contains the proof of Fundamental Theorems and complete
proofs of the Axioms of Infinity, Replacement, and Choice in all
generic extensions. In particular, we were able to fulfill the
promised formal abstract for the Forcing 
Theorems almost to the letter. A requirement for Infinity and the
absoluteness of forcing for atomic formulas, we finished the interface
between our development and
Paulson's constructibility library \cite{paulson_2003} which enables
us to do well-founded
recursion inside transitive models of an appropriate finite fragment
of $\ZF$. As a by-product, we finally met two long-standing goals: the fact that the
generic filter $G$ belongs to the extension $M[G]$ and 
$M\sbq M[G]$. 
In order to take full advantage of the constructibility library
we enhanced it by weakening the assumption of many results and also
extended it with stronger results. % 
Finally, our development is now independent of $\AC$: We modularized
RSL in such a way that a version for countable 
posets does not require choice.

In the course of our work we found it useful to develop Isar methods to
automate repetitive tasks. Part of the interface with Paulson's
library consisted in constructing formulas for each relativized
concept; and actually Isabelle's Simplifier can
synthesize terms for unbound schematic variables in theorems. The
synthesized term, however, is not available outside the theorem; we
introduced a method that creates a definition from a schematic
goal. The second method is concerned with renaming of formulas: we
improved our small library of bounded renamings with a method that
given the source and target environments figures out the renaming
function and produces the relevant lemmas about it.

The source code of our formalization, written for the 2019 version of
Isabelle, can be browsed and downloaded at
\url{https://cs.famaf.unc.edu.ar/~pedro/forcing/}

%% \paragraph{Outline}
%% \label{sec:outline}
%% %% Esto está como subsection para que no parezca como parte de Related
%% %% Work (que tranquilamente podría ir a las conclusiones).
We assume some familiarity with Isabelle and some terminology of set
theory. The current paper is organized as follows. In
Sect.~\ref{sec:isabelle-metatheories} we comment briefly on the
meta-theoretical implications of using Isabelle/ZF. In
Sect.~\ref{sec:relat-absol} we explain the use of relativized concepts
and its importance for the ctm approach. The next sections cover the
core of this report: In Sect.~\ref{sec:definition-forces} we introduce
the definition of the formula transformer $\forceisa$ and reasoning
principles about it; in Sect.~\ref{sec:forcing-theorems} we present
the proofs of the fundamental theorems of forcing. We show in
Sect.~\ref{sec:example-proper-extension} a concrete poset that leads
to a proper extension of the ground model. In
Sect.~\ref{sec:axioms-replacement-choice} we complete the proof that
every axiom and axiom scheme of ZFC is valid in any generic
extension. Sect.~\ref{sec:related-work} briefly discusses related
works and we close the paper by noting the next steps in our project
and drawing conclusions from this formalization.

%%% Local Variables: 
%%% mode: latex
%%% TeX-master: "forcing_in_isabelle_zf"
%%% ispell-local-dictionary: "american"
%%% End: 

\section{Isabelle and (meta)theories}
\label{sec:isabelle-metatheories}

Isabelle \cite{Isabelle,DBLP:books/sp/Paulson94} is a general proof
assistant based on fragment of higher-order logic called
\emph{Pure}. 
The results presented in this work are theorems of a
version of $\ZF$ set theory (without the Axiom of Choice, $\AC$) 
called \emph{Isabelle/ZF}, which is one of the
``object logics'' that can be defined on top of Pure (which is then
used as a language to define rules). Isabelle/ZF defines types
\isatt{i} and \isatt{o} for sets and Booleans, resp., and the $\ZF$
axioms are written down as terms of type \isatt{o}.

More specifically, our results work under the hypothesis of
the existence of a ctm of $\ZFC$.% 
\footnote{By Gödel's Second incompleteness theorem, one must assume at
  least the existence of some model of $\ZF$. 
  The countability is only used to prove the existence of
  generic filters and can be thus replaced in favor of this
  hypothesis.} 
This hypothesis follows, for instance, from the existence of an
inaccessible cardinal. As such, our framework is weaker than those
found usually in type theories with universes, but allows us to work
``Platonistically''--- assuming we are in a universe of sets (namely,
\isatt{i}) and performing constructions there.

On the downside, our approach is not able to provide us with finitary
consistency proofs. It is well known that, for example, the
implication $\Con(\ZF) \implies \Con(\ZFC+\neg\CH)$ can be proved in
\emph{primitive recursive arithmetic (PRA)}. To achieve this, however,
it would have implied to work focusing on the proof mechanisms
and distracting us from our main goal, that is, formalize the ctm
approach currently used by many mathematicians.

It should be noted that Pure is a very weak framework and has no
induction/recursion capabilities. So the only way to define functions
by recursion is inside the object logic. (This works the same for
Isabelle/HOL.) For this reason, to define the relation of forcing, we
needed to resort to \emph{internalized} first-order formulas: they
form a recursively defined set \isatt{formula}. For example, the
predicate of satisfaction
\isatt{sats::i{\isasymRightarrow}i{\isasymRightarrow}i{\isasymRightarrow}o}
(written $M,\mathit{ms}\models\phi$ for a set $M$,
$\mathit{ms}\in\isatt{list}(M)$ and $\phi\in\formula$)
%% (noted \isatt{M, ms \isasymTurnstile{ }\isasymphi} for a set \isatt{M},
%% \isatt{ms {\isasymin} list(M)} and \isatt{{\isasymphi} {\isasymin} formula})
had already been defined by recursion in Paulson~\cite{paulson_2003}.

%%% Local Variables: 
%%% mode: latex
%%% TeX-master: "forcing_in_isabelle_zf"
%%% ispell-local-dictionary: "american"
%%% End: 

\section{Relativization,  absoluteness, and the axioms}
\label{sec:relat-absol}

The concepts of relativization and absoluteness (due to Gödel, in his
proof of the relative consistency of $\AC$ \cite{godel-L}) 
are both
prerequisites and powerful tools in working with transitive
models. A \emph{class} is simply a predicate $C(x)$ with at least one
free variable $x$.
The \emph{relativization} $\phi^C(\bx)$ of a set-theoretic
definition
$\phi$ (of a relation such
as ``$x$ is a subset of $y$'' or of a function like $y=\P(x)$) to
a class $C$ is obtained by restricting all of its quantifiers to $C$.

\[
x \sbq^C y \equiv \forall z.\ C(z) \longrightarrow (z\in x
\longrightarrow z\in y)
\]

The new formula $\phi^C(\bx)$ corresponds to what is obtained by defining
the concept ``inside'' $C$. In fact, for a class corresponding to a
set $c$ (i.e.\ $C(x) \defi x \in c$), the relativization $\phi^C$ of a 
sentence $\phi$ is equivalent to the satisfaction of $\phi$ in the
first-order model $\lb c, \in\rb$.

It turns out that many concepts mean the
same after relativization to a nonempty transitive class $C$; formally
\[
\forall\bx.\ C(\bx) \longrightarrow (\phi^C(\bx) \longleftrightarrow
\phi(\bx))
\]
When this is the case, we say that the relation defined by $\phi$ is
\emph{absolute for transitive models}.\footnote{Absoluteness of
  functions also requires the relativized definition to behave
  functionally over $C$.} As examples, the relation of inclusion
$\subseteq$ ---and actually, any relation defined by a formula
(equivalent to one) using only bounded quantifiers $(\forall x\in y)$
and $(\exists x\in y)$--- is absolute for transitive models. On the
contrary, this is not the case with the powerset operation.

A benefit of working with transitive models
is that many concepts (pairs, unions, and fundamentally ordinals) are
uniform across the universe \isatt{i}, a ctm (of an adequate fragment
of $\ZF$) $M$ and any of its extensions $M[G]$.
For that reason, then one can reason
``externally'' about absolute concepts, instead
of ``inside'' the model; thus, one has at hand any already proved
lemma about the real concept.

Paulson's formalization \cite{paulson_2003} of the relative
consistency of AC by Gödel \cite{godel-L} already contains
absoluteness results which were crucial to our project; we realized
however that they could be refactored to be more useful. The main
objective is to maximize applicability of the relativization machinery
by adjusting the hypothesis of a greater part of its earlier
development. Paulson's architecture had only in mind the consistency
of $\ZFC$, but, for instance, in order to apply it in the development
of forcing, too much is assumed at the beginning; more seriously, some
assumptions cannot be regarded as ``first-order'' (v.g. the
Replacement Scheme) because of the occurrence of predicate variables.
The version we present%
\footnote{%
  While preparing the final version of the present paper,
  our contributions were
  accepted as part of the official Isabelle 2020 release.
}
of the constructibility library,
\isatt{ZF-Constructible-Trans}, weakens the assumptions of many
absoluteness theorems to that of a nonempty transitive class; we also
include some stronger results such as the relativization of powersets.

Apart from the axiom schemes, the $\ZFC$ axioms are initially stated
as predicates on classes (that is, of type
\isatt{(i{\isasymRightarrow}o){\isasymRightarrow}o}); this formulation
allows a better interaction with \isatt{ZF-Constructible}.  The axioms
of Pairing, Union, Foundation, Extensionality, and Infinity are
relativizations of the respective traditional first-order sentences to
the class argument. For the Axiom of Choice we selected a version best
suited for the work with transitive models: the relativization of a
sentence stating that for every $x$ there is surjection from an
ordinal onto $x$. Finally, Separation and Replacement were treated
separately to effectively obtain first-order versions afterwards. It
is to be noted that predicates in Isabelle/ZF are akin to second order
variables and thus do not correspond to first-order formulas.
%% in particular, there is no induction principle for functions
%% of type \isatt{i{\isasymRightarrow}o}. 
For that reason, Separation and Replacement are defined in terms of
\emph{the satisfaction} of an internalized formula $\phi$.  We improved their
specification, with respect to our previous
report~\cite{2019arXiv190103313G}, by lifting the arity restriction
for the parameter $\phi$; consequently we get rid of tupling and thus
various proofs are now slicker.

A benefit of having class versions of the axioms is that we can
apply our synthesis method to obtain their internal, first-order
counterparts. % \footnote{It should be noted that our method for
  % synthetic definition requires that the schematic goal had a
  % particular format where the only unbound schematic variable occurs
  % as an (open) formula whose satisfaction is equivalent to some
  % relativized predicate.}
For the case of the Pairing Axiom, the statement for classes is the
following
\begin{isabelle}
upair{\isacharunderscore}ax{\isacharparenleft}C{\isacharparenright}{\isacharequal}{\isacharequal}{\isasymforall}x{\isacharbrackleft}C{\isacharbrackright}{\isachardot}{\isasymforall}y{\isacharbrackleft}C{\isacharbrackright}{\isachardot}{\isasymexists}z{\isacharbrackleft}C{\isacharbrackright}{\isachardot}\ upair{\isacharparenleft}C{\isacharcomma}x{\isacharcomma}y{\isacharcomma}z{\isacharparenright}
\end{isabelle}
where \isatt{upair} says that \isatt{z} is the unordered pair of
\isatt{x} and \isatt{y}, relative to \isatt{C}.

The following schematic lemma synthesizes its internal version,%
\footnote{The use of such schematic goals and the original definition
  of the collection of lemmas \isatt{sep{\isacharunderscore}rules} are due to Paulson.}
\begin{isabelle}
\isacommand{schematic{\isacharunderscore}goal}\isamarkupfalse%
\ ZF{\isacharunderscore}pairing{\isacharunderscore}auto{\isacharcolon}\isanewline
\ \ \ \ {\isachardoublequoteopen}upair{\isacharunderscore}ax{\isacharparenleft}{\isacharhash}{\isacharhash}A{\isacharparenright}\ {\isasymlongleftrightarrow}\ {\isacharparenleft}A{\isacharcomma}\ {\isacharbrackleft}{\isacharbrackright}\ {\isasymTurnstile}\ {\isacharquery}zfpair{\isacharparenright}{\isachardoublequoteclose}\isanewline
\isacommand{unfolding}\isamarkupfalse%
\ upair{\isacharunderscore}ax{\isacharunderscore}def\ \isanewline
\ \ \isacommand{by}\isamarkupfalse%
\ {\isacharparenleft}{\isacharparenleft}rule\ sep{\isacharunderscore}rules\ {\isacharbar}\ simp{\isacharparenright}{\isacharplus}{\isacharparenright}
\end{isabelle}
and
our \isatt{synthesize} method introduces a new term
\isatt{ZF{\isacharunderscore}pairing{\isacharunderscore}fm} for it:
\begin{isabelle}
\isacommand{synthesize}\isamarkupfalse%
\ {\isachardoublequoteopen}ZF{\isacharunderscore}pairing{\isacharunderscore}fm{\isachardoublequoteclose}\ \isakeyword{from{\isacharunderscore}schematic}\ {\isachardoublequoteopen}ZF{\isacharunderscore}pairing{\isacharunderscore}auto{\isachardoublequoteclose}%
\end{isabelle}
the actual formula obtained is
\isatt{Forall(Forall(Exists(upair{\isacharunderscore}fm(2,1,0))))}.

%%% Local Variables: 
%%% mode: latex
%%% TeX-master: "forcing_in_isabelle_zf"
%%% ispell-local-dictionary: "american"
%%% End: 

\section{The definition of $\forceisa$}
\label{sec:definition-forces}

The core of the development is showing the definability of the
relation of forcing. As we explained in our previous
report~\cite[Sect.~8]{2019arXiv190103313G}, this comprises the
definition of a function $\forceisa$ that maps the set of internal
formulas into itself. It is the very reason of applicability of
forcing that the satisfaction of a first-order formula $\phi$ in all
of the generic extensions of a ctm $M$ can be ``controlled'' in a
definable way from $M$ (viz., by satisfaction of the formula
$\forceisa(\phi)$).

In fact, given a forcing notion $\PP$ (i.e. a preorder with a top element)
in a ctm $M$,
Kunen defines the \emph{forcing relation} model-theoretically 
by considering all extensions $M[G]$ with $G$ generic for $\PP$.
Then two fundamental results are proved, the Truth Lemma and the
Definability Lemma; but the proof of the first one is based on the
formula that witnesses Definability. To make sense of this in our 
formalization, we started with the internalized relation and then
proved that it is equivalent to the semantic version 
(``\isatt{definition{\isacharunderscore}of{\isacharunderscore}forcing},'' in
the next section).
For that reason, the usual notation of the forcing relation 
$p \Vdash \phi\ \mathit{env}$ (for $\mathit{env}$ a list of
``names''), abbreviates in our code the
satisfaction by $M$ of $\forceisa(\phi)$:
\begin{isabelle}
\ \ {\isachardoublequoteopen}p\ {\isasymtturnstile}\ {\isasymphi}\ env\ \ \ {\isasymequiv}\ \ \ M{\isacharcomma}\ {\isacharparenleft}{\isacharbrackleft}p{\isacharcomma}P{\isacharcomma}leq{\isacharcomma}one{\isacharbrackright}\ {\isacharat}\ env{\isacharparenright}
    {\isasymTurnstile}\ forces{\isacharparenleft}{\isasymphi}{\isacharparenright}{\isachardoublequoteclose}
\end{isabelle}

The definition of $\forceisa$ proceeds by recursion
over the set $\formula$ and its base case, that is, for
atomic formulas, is (in)famously the most complicated one. Actually,
newcomers can be puzzled by the fact that forcing for atomic
formulas is also defined by (mutual) recursion: to know if $\tau_1\in\tau_2$ is
forced by $p$ (notation: $\forcesmem(p,\tau_1,\tau_2)$), one must check if $\tau_1=\sigma$ is forced for $\sigma$
moving in the transitive closure of $\tau_2$. To disentangle this, one
must realize that this last recursion must be described syntactically:
the definition of $\forceisa(\phi)$ for atomic $\phi$ is then an
internal definition of the alleged recursion on names. 

Our aim was to follow the definition proposed by Kunen
in~\cite[p.~257]{kunen2011set}, where the following mutual recursion
is given:
\begin{multline}\label{eq:def-forcing-equality}
  \forceseq (p,t_1,t_2) \defi 
  \forall s\in\dom(t_1)\cup\dom(t_2).\ \forall q\pleq p .\\
  \forcesmem(q,s,t_1)\lsii 
  \forcesmem(q,s,t_2),
\end{multline}
\begin{multline}\label{eq:def-forcing-membership}
  \forcesmem(p,t_1,t_2) \defi  \forall v\pleq p. \ \exists q\pleq v. \\
  \exists s.\ \exists r\in \PP .\ \lb s,r\rb \in
      t_2 \land q \pleq r \land \forceseq(q,t_1,s)
\end{multline}
Note that the definition of $\forcesmem$ is equivalent to require 
 the set 
\[
\{q\pleq p : \exists \lb s,r\rb\in t_2 . \ q\pleq r \land \forceseq(q,t_1,s)\}
\]
to be dense below $p$.

It was not straightforward to use the recursion machinery of
Isabelle/ZF to define $\forceseq$ and $\forcesmem$. For this, we
defined a relation $\frecR$ on 4-tuples of elements of $M$, proved
that it is well-founded and, more important, we also proved an
induction principle for this relation:
\begin{isabelle}
\isacommand{lemma}\isamarkupfalse%
\ forces{\isacharunderscore}induction{\isacharcolon}\isanewline
\ \ \isakeyword{assumes}\isanewline
\ \ \ \ {\isachardoublequoteopen}{\isasymAnd}{\isasymtau}\ {\isasymtheta}{\isachardot}\ {\isasymlbrakk}{\isasymAnd}{\isasymsigma}{\isachardot}\ {\isasymsigma}{\isasymin}domain{\isacharparenleft}{\isasymtheta}{\isacharparenright}\ {\isasymLongrightarrow}\ Q{\isacharparenleft}{\isasymtau}{\isacharcomma}{\isasymsigma}{\isacharparenright}{\isasymrbrakk}\ {\isasymLongrightarrow}\ R{\isacharparenleft}{\isasymtau}{\isacharcomma}{\isasymtheta}{\isacharparenright}{\isachardoublequoteclose}\footnotemark\isanewline
\ \ \ \ {\isachardoublequoteopen}{\isasymAnd}{\isasymtau}\ {\isasymtheta}{\isachardot}\ {\isasymlbrakk}{\isasymAnd}{\isasymsigma}{\isachardot}\ {\isasymsigma}{\isasymin}domain{\isacharparenleft}{\isasymtau}{\isacharparenright}\ {\isasymunion}\ domain{\isacharparenleft}{\isasymtheta}{\isacharparenright}\ {\isasymLongrightarrow}\ R{\isacharparenleft}{\isasymsigma}{\isacharcomma}{\isasymtau}{\isacharparenright}\ {\isasymand}\ R{\isacharparenleft}{\isasymsigma}{\isacharcomma}{\isasymtheta}{\isacharparenright}{\isasymrbrakk}\isanewline
\ \ \ \ \ \  {\isasymLongrightarrow}\ Q{\isacharparenleft}{\isasymtau}{\isacharcomma}{\isasymtheta}{\isacharparenright}{\isachardoublequoteclose}\isanewline
\ \ \isakeyword{shows}\isanewline
\ \ \ \ {\isachardoublequoteopen}Q{\isacharparenleft}{\isasymtau}{\isacharcomma}{\isasymtheta}{\isacharparenright}\ {\isasymand}\ R{\isacharparenleft}{\isasymtau}{\isacharcomma}{\isasymtheta}{\isacharparenright}{\isachardoublequoteclose}
\end{isabelle}
\footnotetext{The logical primitives of \emph{Pure} are
\isatt{\isasymLongrightarrow}, \isatt{\&\&\&}, and \isatt{\isasymAnd}
(implication, conjunction, and universal
quantification, resp.), which operate on the meta-Booleans
\isatt{prop}.}
and 
obtained both functions as cases of a another one, 
$\forcesat$, using a single recursion on $\frecR$. Then we obtained 
(\ref{eq:def-forcing-equality}) and (\ref{eq:def-forcing-membership})
as our corollaries \isatt{def{\isacharunderscore}forces{\isacharunderscore}eq} and
\isatt{def{\isacharunderscore}forces{\isacharunderscore}mem}.

Other approaches, like the one in Neeman~\cite{neeman-course} (and
Kunen's older book \cite{kunen1980}), prefer
to have a single, more complicated, definition by simple recursion for
$\forceseq$ and then define $\forcesmem$ explicitly. On hindsight,
this might have been a little simpler to do, but we preferred to be as
faithful to the text as possible concerning this point.

Once $\forcesat$ and its relativized version
$\isatt{is{\isacharunderscore}forces{\isacharunderscore}at}$ were
defined, we proceeded to show absoluteness and provided internal
definitions for the recursion on names using results in
\isatt{ZF-Constructible}. This finished the atomic case of the
formula-transformer $\forceisa$. The characterization of $\forceisa$
for negated and universal quantified formulas is given by the
following lemmas, respectively:
\begin{isabelle}
\isacommand{lemma}\isamarkupfalse%
\ sats{\isacharunderscore}forces{\isacharunderscore}Neg{\isacharcolon}\isanewline
\ \ \isakeyword{assumes}\isanewline
\ \ \ \ {\isachardoublequoteopen}p{\isasymin}P{\isachardoublequoteclose}\ {\isachardoublequoteopen}env\ {\isasymin}\ list{\isacharparenleft}M{\isacharparenright}{\isachardoublequoteclose}\ {\isachardoublequoteopen}{\isasymphi}{\isasymin}formula{\isachardoublequoteclose}\isanewline
\ \ \isakeyword{shows}\isanewline
\ \ \ \ {\isachardoublequoteopen}M{\isacharcomma}\ {\isacharbrackleft}p{\isacharcomma}P{\isacharcomma}leq{\isacharcomma}one{\isacharbrackright}\ {\isacharat}\ env\ {\isasymTurnstile}\ forces{\isacharparenleft}Neg{\isacharparenleft}{\isasymphi}{\isacharparenright}{\isacharparenright}\ \ \ {\isasymlongleftrightarrow}\ \isanewline
\ \ \ \ \ {\isasymnot}{\isacharparenleft}{\isasymexists}q{\isasymin}M{\isachardot}\ q{\isasymin}P\ {\isasymand}\ is{\isacharunderscore}leq{\isacharparenleft}{\isacharhash}{\isacharhash}M{\isacharcomma}leq{\isacharcomma}q{\isacharcomma}p{\isacharparenright}\ {\isasymand}\ \isanewline
\ \ \ \ \ \ \ \ \ \ M{\isacharcomma}\ {\isacharbrackleft}q{\isacharcomma}P{\isacharcomma}leq{\isacharcomma}one{\isacharbrackright}{\isacharat}env\ {\isasymTurnstile}\ forces{\isacharparenleft}{\isasymphi}{\isacharparenright}{\isacharparenright}{\isachardoublequoteclose}\isanewline

\isacommand{lemma}\isamarkupfalse%
\ sats{\isacharunderscore}forces{\isacharunderscore}Forall{\isacharcolon}\isanewline
\ \ \isakeyword{assumes}\isanewline
\ \ \ \ {\isachardoublequoteopen}p{\isasymin}P{\isachardoublequoteclose}\ {\isachardoublequoteopen}env\ {\isasymin}\ list{\isacharparenleft}M{\isacharparenright}{\isachardoublequoteclose}\ {\isachardoublequoteopen}{\isasymphi}{\isasymin}formula{\isachardoublequoteclose}\isanewline
\ \ \isakeyword{shows}\isanewline
\ \ \ \ {\isachardoublequoteopen}M{\isacharcomma}{\isacharbrackleft}p{\isacharcomma}P{\isacharcomma}leq{\isacharcomma}one{\isacharbrackright}\ {\isacharat}\ env\ {\isasymTurnstile}\ forces{\isacharparenleft}Forall{\isacharparenleft}{\isasymphi}{\isacharparenright}{\isacharparenright}\ {\isasymlongleftrightarrow}\ \isanewline
\ \ \ \ \ {\isacharparenleft}{\isasymforall}x{\isasymin}M{\isachardot}\ \ \ M{\isacharcomma}\ {\isacharbrackleft}p{\isacharcomma}P{\isacharcomma}leq{\isacharcomma}one{\isacharcomma}x{\isacharbrackright}\ {\isacharat}\ env\ {\isasymTurnstile}\ forces{\isacharparenleft}{\isasymphi}{\isacharparenright}{\isacharparenright}{\isachardoublequoteclose}
\end{isabelle}

Let us note in passing another improvement over our previous report:
we made a couple of new technical results concerning recursive
definitions. Paulson proved absoluteness of functions defined by
well-founded recursion over a transitive relation. Some of our
definitions by recursion (\emph{check} and \emph{forces}) do not fit
in that scheme.  One can replace the relation $R$ for its transitive
closure $R^+$ in the recursive definition because one can prove, in
general, that
$F\!\upharpoonright\!(R^{-1}(x))(y) =
F\!\upharpoonright\!\bigl((R^+)^{-1}(x)\bigr)(y)$ whenever $(x,y) \in R$.

%%% Local Variables: 
%%% mode: latex
%%% TeX-master: "forcing_in_isabelle_zf"
%%% ispell-local-dictionary: "american"
%%% End: 

\section{The forcing theorems}
\label{sec:forcing-theorems}

After the definition of $\forceisa$ is complete, the proof of the
Fundamental Theorems of Forcing is comparatively straightforward, and
we were able to follow Kunen very closely. The more involved points of
this part of the development were those where we needed to prove that
various (dense) subsets of $\PP$ were in $M$; for this, we have
resorted to several ad-hoc absoluteness lemmas.

The first results concern characterizations of the forcing
relation. Two of them are \isatt{Forces{\isacharunderscore}Member}:
\begin{center}
  \isatt{{\isacharparenleft}p\ {\isasymtturnstile}\ Member{\isacharparenleft}n{\isacharcomma}m{\isacharparenright}\ env{\isacharparenright}\ {\isasymlongleftrightarrow}\ forces{\isacharunderscore}mem{\isacharparenleft}p{\isacharcomma}t{\isadigit{1}}{\isacharcomma}t{\isadigit{2}}{\isacharparenright}},
\end{center}
where \isatt{t{\isadigit{1}}} and \isatt{t{\isadigit{1}}} are the
\isatt{n}th resp.\ \isatt{m}th elements of \isatt{env}, and  \isatt{Forces{\isacharunderscore}Forall}:
\begin{center}
  \isatt{{\isacharparenleft}p\ {\isasymtturnstile}\ Forall{\isacharparenleft}{\isasymphi}{\isacharparenright}\ env{\isacharparenright}\ {\isasymlongleftrightarrow}\ {\isacharparenleft}{\isasymforall}x{\isasymin}M{\isachardot}\ {\isacharparenleft}p\ {\isasymtturnstile}\ {\isasymphi}\ {\isacharparenleft}{\isacharbrackleft}x{\isacharbrackright}\ {\isacharat}\ env{\isacharparenright}{\isacharparenright}{\isacharparenright}}.
\end{center}
Equivalent statements, along with the ones corresponding to \isatt{Forces{\isacharunderscore}Equal} and
\isatt{Forces{\isacharunderscore}Nand}, appear in Kunen as the
inductive definition of the forcing relation \cite[Def.~IV.2.42]{kunen2011set}.

As with the previous section, the proofs of the forcing theorems have two different
flavors: The ones for the atomic formulas proceed by using the
principle of 
\isatt{forces{\isacharunderscore}induction}, and then an induction on
$\formula$ wraps the former with the remaining cases (\isatt{Nand} and \isatt{Forall}). 

As an example of the first class, we can take a look at our
formalization of \cite[Lem.~IV.2.40(a)]{kunen2011set}. Note that the
context (a ``locale,'' in Isabelle terminology, namely \isatt{forcing{\isacharunderscore}data}) of the lemma 
includes the assumption of \isatt{P} being
a forcing notion, and the predicate of being $M$-generic is defined in
terms of \isatt{P}:

\begin{isabelle}
  \isacommand{lemma}\isamarkupfalse%
  \ IV{\isadigit{2}}{\isadigit{4}}{\isadigit{0}}a{\isacharcolon}\isanewline
  \ \ \isakeyword{assumes}\isanewline
  \ \ \ \ {\isachardoublequoteopen}M{\isacharunderscore}generic{\isacharparenleft}G{\isacharparenright}{\isachardoublequoteclose}\isanewline
  \ \ \isakeyword{shows}\ \isanewline
  \ \ {\isachardoublequoteopen}{\isacharparenleft}{\isasymtau}{\isasymin}M{\isasymlongrightarrow}{\isasymtheta}{\isasymin}M{\isasymlongrightarrow}{\isacharparenleft}{\isasymforall}p{\isasymin}G{\isachardot}forces{\isacharunderscore}eq{\isacharparenleft}p{\isacharcomma}{\isasymtau}{\isacharcomma}{\isasymtheta}{\isacharparenright}{\isasymlongrightarrow}val{\isacharparenleft}G{\isacharcomma}{\isasymtau}{\isacharparenright}{\isacharequal}val{\isacharparenleft}G{\isacharcomma}{\isasymtheta}{\isacharparenright}{\isacharparenright}{\isacharparenright}\isanewline
  \ \  {\isasymand}\isanewline
  \ \ {\isacharparenleft}{\isasymtau}{\isasymin}M{\isasymlongrightarrow}{\isasymtheta}{\isasymin}M{\isasymlongrightarrow}{\isacharparenleft}{\isasymforall}p{\isasymin}G{\isachardot}forces{\isacharunderscore}mem{\isacharparenleft}p{\isacharcomma}{\isasymtau}{\isacharcomma}{\isasymtheta}{\isacharparenright}{\isasymlongrightarrow}val{\isacharparenleft}G{\isacharcomma}{\isasymtau}{\isacharparenright}{\isasymin}val{\isacharparenleft}G{\isacharcomma}{\isasymtheta}{\isacharparenright}{\isacharparenright}{\isacharparenright}{\isachardoublequoteclose}
\end{isabelle}
Its proof starts by an introduction of \isatt{forces{\isacharunderscore}induction};
the  inductive cases for each atomic type were handled before as
separate lemmas (\isatt{IV240a{\isacharunderscore}mem} and \isatt{IV240a{\isacharunderscore}eq}). We
illustrate with the statement of the latter.
\begin{isabelle}
\isacommand{lemma}\isamarkupfalse%
\ IV{\isadigit{2}}{\isadigit{4}}{\isadigit{0}}a{\isacharunderscore}eq{\isacharcolon}\isanewline
\ \ \isakeyword{assumes}\isanewline
\ \ \ \ {\isachardoublequoteopen}M{\isacharunderscore}generic{\isacharparenleft}G{\isacharparenright}{\isachardoublequoteclose}\ {\isachardoublequoteopen}p{\isasymin}G{\isachardoublequoteclose}\ {\isachardoublequoteopen}forces{\isacharunderscore}eq{\isacharparenleft}p{\isacharcomma}{\isasymtau}{\isacharcomma}{\isasymtheta}{\isacharparenright}{\isachardoublequoteclose}\isanewline
\ \ \ \ \isakeyword{and}\isanewline
\ \ \ \ IH{\isacharcolon}{\isachardoublequoteopen}{\isasymAnd}q\ {\isasymsigma}{\isachardot}\ q{\isasymin}P\ {\isasymLongrightarrow}\ q{\isasymin}G\ {\isasymLongrightarrow}\ {\isasymsigma}{\isasymin}domain{\isacharparenleft}{\isasymtau}{\isacharparenright}\ {\isasymunion}\ domain{\isacharparenleft}{\isasymtheta}{\isacharparenright}\ {\isasymLongrightarrow}\ \isanewline
\ \ \ \ \ \ \ \ {\isacharparenleft}forces{\isacharunderscore}mem{\isacharparenleft}q{\isacharcomma}{\isasymsigma}{\isacharcomma}{\isasymtau}{\isacharparenright}\ {\isasymlongrightarrow}\ val{\isacharparenleft}G{\isacharcomma}{\isasymsigma}{\isacharparenright}\ {\isasymin}\ val{\isacharparenleft}G{\isacharcomma}{\isasymtau}{\isacharparenright}{\isacharparenright}\ {\isasymand}\isanewline
\ \ \ \ \ \ \ \ {\isacharparenleft}forces{\isacharunderscore}mem{\isacharparenleft}q{\isacharcomma}{\isasymsigma}{\isacharcomma}{\isasymtheta}{\isacharparenright}\ {\isasymlongrightarrow}\ val{\isacharparenleft}G{\isacharcomma}{\isasymsigma}{\isacharparenright}\ {\isasymin}\ val{\isacharparenleft}G{\isacharcomma}{\isasymtheta}{\isacharparenright}{\isacharparenright}{\isachardoublequoteclose}\isanewline
\ \ \isakeyword{shows}\isanewline
\ \ \ \ {\isachardoublequoteopen}val{\isacharparenleft}G{\isacharcomma}{\isasymtau}{\isacharparenright}\ {\isacharequal}\ val{\isacharparenleft}G{\isacharcomma}{\isasymtheta}{\isacharparenright}{\isachardoublequoteclose}
\end{isabelle}

Examples of proofs  using the second kind of induction include
the basic \isatt{strengthening{\isacharunderscore}lemma} and the main
results in this section, the lemmas of  Density  (actually, its nontrivial
direction
\isatt{dense{\isacharunderscore}below{\isacharunderscore}imp{\isacharunderscore}forces})
and Truth, 
which we state next.
\begin{isabelle}
\isacommand{lemma}\isamarkupfalse%
\ density{\isacharunderscore}lemma{\isacharcolon}\isanewline
\ \ \isakeyword{assumes}\isanewline
\ \ \ \ {\isachardoublequoteopen}p{\isasymin}P{\isachardoublequoteclose}\ {\isachardoublequoteopen}{\isasymphi}{\isasymin}formula{\isachardoublequoteclose}\ {\isachardoublequoteopen}env{\isasymin}list{\isacharparenleft}M{\isacharparenright}{\isachardoublequoteclose}\ {\isachardoublequoteopen}arity{\isacharparenleft}{\isasymphi}{\isacharparenright}{\isasymle}length{\isacharparenleft}env{\isacharparenright}{\isachardoublequoteclose}\isanewline
\ \ \isakeyword{shows}\isanewline
\ \ \ \ {\isachardoublequoteopen}{\isacharparenleft}p\ {\isasymtturnstile}\ {\isasymphi}\ env{\isacharparenright}\ {\isasymlongleftrightarrow}\ dense{\isacharunderscore}below{\isacharparenleft}{\isacharbraceleft}q{\isasymin}P{\isachardot}\ {\isacharparenleft}q\ {\isasymtturnstile}\ {\isasymphi}\ env{\isacharparenright}{\isacharbraceright}{\isacharcomma}p{\isacharparenright}{\isachardoublequoteclose}
\end{isabelle}
\begin{isabelle}
\isacommand{lemma}\isamarkupfalse%
\ truth{\isacharunderscore}lemma{\isacharcolon}\isanewline
\ \ \isakeyword{assumes}\ \isanewline
\ \ \ \ {\isachardoublequoteopen}{\isasymphi}{\isasymin}formula{\isachardoublequoteclose}\ {\isachardoublequoteopen}M{\isacharunderscore}generic{\isacharparenleft}G{\isacharparenright}{\isachardoublequoteclose}\isanewline
\ \ \isakeyword{shows}\ \isanewline
\ \ \ \ \ {\isachardoublequoteopen}{\isasymAnd}env{\isachardot}\ env{\isasymin}list{\isacharparenleft}M{\isacharparenright}\ {\isasymLongrightarrow}\ arity{\isacharparenleft}{\isasymphi}{\isacharparenright}{\isasymle}length{\isacharparenleft}env{\isacharparenright}\ {\isasymLongrightarrow}\ \isanewline
\ \ \ \ \ \ {\isacharparenleft}{\isasymexists}p{\isasymin}G{\isachardot}\ {\isacharparenleft}p\ {\isasymtturnstile}\ {\isasymphi}\ env{\isacharparenright}{\isacharparenright}\ \ {\isasymlongleftrightarrow}\ \ M{\isacharbrackleft}G{\isacharbrackright}{\isacharcomma}\ map{\isacharparenleft}val{\isacharparenleft}G{\isacharparenright}{\isacharcomma}env{\isacharparenright}\ {\isasymTurnstile}\ {\isasymphi}{\isachardoublequoteclose}
\end{isabelle}
From these results, the semantical characterization of the forcing
relation (the ``definition of $\forces$'' in 
\cite[IV.2.22]{kunen2011set}) follows easily:
\begin{isabelle}
\isacommand{lemma}\isamarkupfalse%
\ definition{\isacharunderscore}of{\isacharunderscore}forcing{\isacharcolon}\isanewline
\ \ \isakeyword{assumes}\isanewline
\ \ \ \ {\isachardoublequoteopen}p{\isasymin}P{\isachardoublequoteclose}\ {\isachardoublequoteopen}{\isasymphi}{\isasymin}formula{\isachardoublequoteclose}\ {\isachardoublequoteopen}env{\isasymin}list{\isacharparenleft}M{\isacharparenright}{\isachardoublequoteclose}\ {\isachardoublequoteopen}arity{\isacharparenleft}{\isasymphi}{\isacharparenright}{\isasymle}length{\isacharparenleft}env{\isacharparenright}{\isachardoublequoteclose}\isanewline
\ \ \isakeyword{shows}\isanewline
\ \ \ \ {\isachardoublequoteopen}{\isacharparenleft}p\ {\isasymtturnstile}\ {\isasymphi}\ env{\isacharparenright}\ {\isasymlongleftrightarrow}\isanewline
\ \ \ \ \ {\isacharparenleft}{\isasymforall}G{\isachardot}\ M{\isacharunderscore}generic{\isacharparenleft}G{\isacharparenright}{\isasymand}\ p{\isasymin}G\ {\isasymlongrightarrow}\ M{\isacharbrackleft}G{\isacharbrackright}{\isacharcomma}\ map{\isacharparenleft}val{\isacharparenleft}G{\isacharparenright}{\isacharcomma}env{\isacharparenright}\ {\isasymTurnstile}\ {\isasymphi}{\isacharparenright}{\isachardoublequoteclose}
\end{isabelle}

The present statement of the Fundamental Theorems is almost exactly
the same of those in our previous report \cite{2019arXiv190103313G},
with the only modification being the bound on arities and a missing
typing constraint. This implied only minor adjustments in the proofs
of the satisfaction of axioms.

%%% Local Variables: 
%%% mode: latex
%%% TeX-master: "forcing_in_isabelle_zf"
%%% ispell-local-dictionary: "american"
%%% End: 

\section{Example of proper extension}
\label{sec:example-proper-extension}

Even when the axioms of $\ZFC$ are proved in the generic extension,
one cannot claim that the magic of forcing has taken place unless one
is able to provide some \emph{proper} extension with the \emph{same
ordinals}. After all, one is assuming from the start a model $M$ of $\ZFC$,
and in some trivial cases $M[G]$ might end up to be exactly $M$; this
is where \emph{proper} enters the stage. But, for instance, in the
presence of large cardinals, a model $M'\supsetneq M$ might be an
end-extension of $M$ ---this is where we ask the two models to have the
same ordinals, the same \emph{height}. 

Three theory files contain the relevant
results. \verb|Ordinals_In_MG.thy| shows, using the closure of $M$
under ranks, that $M$ and $M[G]$ share the same ordinals (actually,
ranks of elements of $M[G]$ are bounded by the ranks of their names in
$M$):
\begin{isabelle}
\isacommand{lemma}\isamarkupfalse%
\ rank{\isacharunderscore}val{\isacharcolon}\ {\isachardoublequoteopen}rank{\isacharparenleft}val{\isacharparenleft}G{\isacharcomma}x{\isacharparenright}{\isacharparenright}\ {\isasymle}\ rank{\isacharparenleft}x{\isacharparenright}{\isachardoublequoteclose}\isanewline
\isacommand{lemma}\isamarkupfalse%
\ Ord{\isacharunderscore}MG{\isacharunderscore}iff{\isacharcolon}\isanewline
\ \ \isakeyword{assumes}\ {\isachardoublequoteopen}Ord{\isacharparenleft}{\isasymalpha}{\isacharparenright}{\isachardoublequoteclose}\ \isanewline
\ \ \isakeyword{shows}\ {\isachardoublequoteopen}{\isasymalpha}\ {\isasymin}\ M\ {\isasymlongleftrightarrow}\ {\isasymalpha}\ {\isasymin}\ M{\isacharbrackleft}G{\isacharbrackright}{\isachardoublequoteclose}
\end{isabelle}
To prove these results, we found it useful to formalize induction over
the relation \isatt{ed}$(x,y) \defi x\in\dom(y)$, which is key
to arguments involving names.

\verb|Succession_Poset.thy| contains our first example of a poset
that interprets the locale
\isatt{forcing{\isacharunderscore}notion}, essentially the notion for
adding one Cohen real. It is the set $2^{<\om}$ of all finite binary
sequences partially  ordered by reverse inclusion.
The sufficient condition for a proper extension is that
the forcing poset is \emph{separative}: every element has two
incompatible (\isatt{{\isasymbottom}s}) extensions. Here,
\isatt{seq{\isacharunderscore}upd{\isacharparenleft}f{\isacharcomma}x{\isacharparenright}}
adds \isatt{x} to the end of the sequence \isatt{f}.

\begin{isabelle}
\isacommand{lemma}\isamarkupfalse%
\ seqspace{\isacharunderscore}separative{\isacharcolon}\isanewline
\ \ \isakeyword{assumes}\ {\isachardoublequoteopen}f{\isasymin}{\isadigit{2}}{\isacharcircum}{\isacharless}{\isasymomega}{\isachardoublequoteclose}\isanewline
\ \ \isakeyword{shows}\ {\isachardoublequoteopen}seq{\isacharunderscore}upd{\isacharparenleft}f{\isacharcomma}{\isadigit{0}}{\isacharparenright}\ {\isasymbottom}s\ seq{\isacharunderscore}upd{\isacharparenleft}f{\isacharcomma}{\isadigit{1}}{\isacharparenright}{\isachardoublequoteclose}
\end{isabelle}
 
We prove in the theory file \verb|Proper_Extension.thy| that, in
general, every separative forcing notion gives rise to a proper
extension.

%%% Local Variables: 
%%% mode: latex
%%% TeX-master: "forcing_in_isabelle_zf"
%%% ispell-local-dictionary: "american"
%%% End: 

\section{The axioms of replacement and choice}
\label{sec:axioms-replacement-choice}

In our report~\cite{2019arXiv190103313G} we proved that any generic
extension preserves the satisfaction of almost all the axioms,
including the separation scheme (we adapted those proofs to the
current statement of the axiom schemes). Our proofs that Replacement
and choice hold in every generic extension depend on further
relativized concepts and closure properties.

\subsection{Replacement}

The proof of the Replacement Axiom scheme in $M[G]$ in Kunen uses the
Reflection Principle relativized to $M$. We took an alternative
pathway, following Neeman \cite{neeman-course}. In his course notes,
he uses the relativization of the cumulative hierarchy of sets. 

The
family of all sets of rank less than $\alpha$ is called
\isatt{Vset}$(\alpha)$ in Isabelle/ZF. We showed, in the theory file
\verb|Relative_Univ.thy|
 the following
relativization and closure results concerning this function, for a
class $M$ satisfying the locale \isatt{M{\isacharunderscore}eclose}
plus the Powerset Axiom and four instances of replacement.
\begin{isabelle}
\isacommand{lemma}\isamarkupfalse%
\ Vset{\isacharunderscore}abs{\isacharcolon}\ {\isachardoublequoteopen}{\isasymlbrakk}\ M{\isacharparenleft}i{\isacharparenright}{\isacharsemicolon}\ M{\isacharparenleft}V{\isacharparenright}{\isacharsemicolon}\ Ord{\isacharparenleft}i{\isacharparenright}\ {\isasymrbrakk}\ {\isasymLongrightarrow}\ \isanewline
\ \ \ \ \ \ \ \  \ \  \ \ \ \ \ \ \ \ \ \ is{\isacharunderscore}Vset{\isacharparenleft}M{\isacharcomma}i{\isacharcomma}V{\isacharparenright}\ {\isasymlongleftrightarrow}\ V\ {\isacharequal}\ {\isacharbraceleft}x{\isasymin}Vset{\isacharparenleft}i{\isacharparenright}{\isachardot}\ M{\isacharparenleft}x{\isacharparenright}{\isacharbraceright}{\isachardoublequoteclose}
\end{isabelle}
\begin{isabelle}
\isacommand{lemma}\isamarkupfalse%
\ Vset{\isacharunderscore}closed{\isacharcolon}\ {\isachardoublequoteopen}{\isasymlbrakk}\ M{\isacharparenleft}i{\isacharparenright}{\isacharsemicolon}\ Ord{\isacharparenleft}i{\isacharparenright}\ {\isasymrbrakk}\ {\isasymLongrightarrow}\ M{\isacharparenleft}{\isacharbraceleft}x{\isasymin}Vset{\isacharparenleft}i{\isacharparenright}{\isachardot}\ M{\isacharparenleft}x{\isacharparenright}{\isacharbraceright}{\isacharparenright}{\isachardoublequoteclose}
\end{isabelle}
We also have the basic result
\begin{isabelle}
\isacommand{lemma}\isamarkupfalse%
\ M{\isacharunderscore}into{\isacharunderscore}Vset{\isacharcolon}\isanewline
\ \ \isakeyword{assumes}\ {\isachardoublequoteopen}M{\isacharparenleft}a{\isacharparenright}{\isachardoublequoteclose}\isanewline
\ \ \isakeyword{shows}\ {\isachardoublequoteopen}{\isasymexists}i{\isacharbrackleft}M{\isacharbrackright}{\isachardot}\ {\isasymexists}V{\isacharbrackleft}M{\isacharbrackright}{\isachardot}\ ordinal{\isacharparenleft}M{\isacharcomma}i{\isacharparenright}\ {\isasymand}\ is{\isacharunderscore}Vfrom{\isacharparenleft}M{\isacharcomma}{\isadigit{0}}{\isacharcomma}i{\isacharcomma}V{\isacharparenright}\ {\isasymand}\ a{\isasymin}V{\isachardoublequoteclose}
\end{isabelle}
stating that $M$ is included in 
$\union\{\isatt{Vset}^M(\alpha) : \alpha\in M\}$ (actually they are equal).

For the proof of the Replacement Axiom, we assume that $\phi$ is
functional in its first two variables when interpreted in $M[G]$ and
the first ranges over the domain \isatt{c}${}\in M[G]$. Then we show
that the collection of
all values of the second variable, when the first ranges over
\isatt{c}, belongs to $M[G]$:
\begin{isabelle}
\isacommand{lemma}\isamarkupfalse%
\ Replace{\isacharunderscore}sats{\isacharunderscore}in{\isacharunderscore}MG{\isacharcolon}\isanewline
\ \ \isakeyword{assumes}\isanewline
\ \ \ \ {\isachardoublequoteopen}c{\isasymin}M{\isacharbrackleft}G{\isacharbrackright}{\isachardoublequoteclose}\ {\isachardoublequoteopen}env\ {\isasymin}\ list{\isacharparenleft}M{\isacharbrackleft}G{\isacharbrackright}{\isacharparenright}{\isachardoublequoteclose}\isanewline
\ \ \ \ {\isachardoublequoteopen}{\isasymphi}\ {\isasymin}\ formula{\isachardoublequoteclose}\ {\isachardoublequoteopen}arity{\isacharparenleft}{\isasymphi}{\isacharparenright}\ {\isasymle}\ {\isadigit{2}}\ {\isacharhash}{\isacharplus}\ length{\isacharparenleft}env{\isacharparenright}{\isachardoublequoteclose}\isanewline
\ \ \ \ {\isachardoublequoteopen}univalent{\isacharparenleft}{\isacharhash}{\isacharhash}M{\isacharbrackleft}G{\isacharbrackright}{\isacharcomma}\ c{\isacharcomma}\ {\isasymlambda}x\ v{\isachardot}\ {\isacharparenleft}M{\isacharbrackleft}G{\isacharbrackright}{\isacharcomma}\ {\isacharbrackleft}x{\isacharcomma}v{\isacharbrackright}{\isacharat}env\ {\isasymTurnstile}\ {\isasymphi}{\isacharparenright}{\isacharparenright}{\isachardoublequoteclose}\isanewline
\ \ \isakeyword{shows}\isanewline
\ \ \ \ {\isachardoublequoteopen}{\isacharbraceleft}v{\isachardot}\ x{\isasymin}c{\isacharcomma}\ v{\isasymin}M{\isacharbrackleft}G{\isacharbrackright}\ {\isasymand}\ {\isacharparenleft}M{\isacharbrackleft}G{\isacharbrackright}{\isacharcomma}\ {\isacharbrackleft}x{\isacharcomma}v{\isacharbrackright}{\isacharat}env\ {\isasymTurnstile}\ {\isasymphi}{\isacharparenright}{\isacharbraceright}\ {\isasymin}\ M{\isacharbrackleft}G{\isacharbrackright}{\isachardoublequoteclose}
\end{isabelle}
From this, the satisfaction of the Replacement Axiom in $M[G]$ follows
very easily.

The proof of the previous lemma, following Neeman, proceeds as usual
by turning an argument concerning elements of $M[G]$ to one involving
names lying in $M$, and connecting both worlds by using the forcing
theorems. In the case at hand, by functionality of $\phi$ we know that
for every $x\in c\cap M[G]$ there exists exactly one $v\in M[G]$ such
that
$M[G], [x,v]\mathbin{@} \mathit{env} \models \phi$. Now,
given a name $\pi'\in M$ for $c$, every name of an element of $c$
belongs to $\pi\defi \dom(\pi')\times \PP$, which is easily seen to be
in $M$. We will use $\pi$ to be the domain in an application of the
Replacement Axiom in $M$. But now, obviously, we have lost
functionality since there are many names $\dot v\in M$ for a fixed $v$
in $M[G]$. To solve this issue, for each $\rho p \defi\lb\rho,p\rb\in
\pi$ we calculate the
minimum rank of some $\tau\in M$ such that 
$p\forces \phi(\rho,\tau,\dots)$ if there is one, or $0$ otherwise. By
Replacement in $M$, we can show that the supremum \isatt{?sup} of these ordinals
belongs to $M$ and we can construct a \isatt{?bigname} $\defi$ 
\isatt{{\isacharbraceleft}x{\isasymin}Vset{\isacharparenleft}{\isacharquery}sup{\isacharparenright}{\isachardot}\ x\ {\isasymin}\
}$M$\isatt{{\isacharbraceright}\ {\isasymtimes}\ {\isacharbraceleft}one{\isacharbraceright}}
whose interpretation by (any generic) $G$ will include all possible elements
as $v$ above.

The previous calculation required some absoluteness and closure
results regarding the minimum ordinal binder, \isatt{Least}$(Q)$, also
denoted $\mu x. Q(x)$, that can be found in the theory file
\verb|Least.thy|.

\subsection{Choice}
A first important observation is that the proof of $\AC$ in $M[G]$
only requires the assumption that $M$ satisfies (a finite fragment of)
$\ZFC$. There is no need to invoke Choice in the metatheory.

Although our previous version of the development used $\AC$, that was
only needed to show the Rasiowa-Sikorski Lemma (RSL) for
arbitrary posets. We have modularized the proof of the latter
and now the version for countable posets that we use to show the
existence of generic filters
does not require Choice (as it is well known). We also bundled the
full RSL along with our implementation of the principle of dependent
choices in an independent branch of the dependency graph, which is the
only place where the theory \texttt{ZF.AC} is invoked.

Our statement of the Axiom of Choice is the one preferred for
arguments involving transitive classes satisfying $\ZF$:
\begin{center}
\isatt{{\isasymforall}x{\isacharbrackleft}M{\isacharbrackright}{\isachardot}\ {\isasymexists}a{\isacharbrackleft}M{\isacharbrackright}{\isachardot}\ {\isasymexists}f{\isacharbrackleft}M{\isacharbrackright}{\isachardot}\ ordinal{\isacharparenleft}M{\isacharcomma}a{\isacharparenright}\ {\isasymand}\ surjection{\isacharparenleft}M{\isacharcomma}a{\isacharcomma}x{\isacharcomma}f{\isacharparenright}}
\end{center}
The Simplifier is able to show automatically that this
statement is equivalent to the next one, in which the real notions of
ordinal and surjection appear:
\begin{center}
\isatt{{\isasymforall}x{\isacharbrackleft}M{\isacharbrackright}{\isachardot}\ {\isasymexists}a{\isacharbrackleft}M{\isacharbrackright}{\isachardot}\ {\isasymexists}f{\isacharbrackleft}M{\isacharbrackright}{\isachardot}\ Ord{\isacharparenleft}a{\isacharparenright}\ {\isasymand}\ f\ {\isasymin}\ surj{\isacharparenleft}a{\isacharcomma}x{\isacharparenright}}
\end{center}

As with the forcing axioms, the proof of $\AC$ in $M[G]$ follows the pattern of Kunen
\cite[IV.2.27]{kunen2011set} and is rather
straightforward; the only complicated technical point being to show
that the relevant name belongs to $M$. We assume that \isatt{a}${}\neq\emptyset$
belongs to $M[G]$ and has a name $\tau\in M$. By $\AC$ in $M$, there
is a surjection \isatt{s} from an ordinal $\alpha\in M$ ($\subseteq M[G]$) onto
$\dom(\tau)$. Now
\begin{center}
\isatt{{\isacharbraceleft}opair{\isacharunderscore}name{\isacharparenleft}check{\isacharparenleft}{\isasymbeta}{\isacharparenright}{\isacharcomma}s{\isacharbackquote}{\isasymbeta}{\isacharparenright}{\isachardot}\ {\isasymbeta}{\isasymin}{\isasymalpha}{\isacharbraceright}\ {\isasymtimes}\ {\isacharbraceleft}one{\isacharbraceright}}
\end{center}
is a name for a function \isatt{f} with domain $\alpha$ such that \isatt{a}
is included in its range, and where
\isatt{opair{\isacharunderscore}name}$(\sig,\rho)$ is a name for the
ordered pair $\lb\val(G,\sig),\val(G,\rho)\rb$. From this, $\AC$ in
$M[G]$ follows easily.

\subsection{The main theorem}
With all these elements in place, we are able to transcript the main
theorem of our formalization:
\begin{isabelle}
\isacommand{theorem}\isamarkupfalse%
\ extensions{\isacharunderscore}of{\isacharunderscore}ctms{\isacharcolon}\isanewline
\ \ \isakeyword{assumes}\ \isanewline
\ \ \ \ {\isachardoublequoteopen}M\ {\isasymapprox}\ nat{\isachardoublequoteclose}\ {\isachardoublequoteopen}Transset{\isacharparenleft}M{\isacharparenright}{\isachardoublequoteclose}\ {\isachardoublequoteopen}M\ {\isasymTurnstile}\ ZF{\isachardoublequoteclose}\isanewline
\ \ \isakeyword{shows}\ \isanewline
\ \ \ \ {\isachardoublequoteopen}{\isasymexists}N{\isachardot}\ \isanewline
\ \ \ \ \ \ M\ {\isasymsubseteq}\ N\ {\isasymand}\ N\ {\isasymapprox}\ nat\ {\isasymand}\ Transset{\isacharparenleft}N{\isacharparenright}\ {\isasymand}\ N\ {\isasymTurnstile}\ ZF\ {\isasymand}\ M{\isasymnoteq}N\ {\isasymand}\isanewline
\ \ \ \ \ \ {\isacharparenleft}{\isasymforall}{\isasymalpha}{\isachardot}\ Ord{\isacharparenleft}{\isasymalpha}{\isacharparenright}\ {\isasymlongrightarrow}\ {\isacharparenleft}{\isasymalpha}\ {\isasymin}\ M\ {\isasymlongleftrightarrow}\ {\isasymalpha}\ {\isasymin}\ N{\isacharparenright}{\isacharparenright}\ {\isasymand}\isanewline
\ \ \ \ \ \ {\isacharparenleft}M{\isacharcomma}\ {\isacharbrackleft}{\isacharbrackright}{\isasymTurnstile}\ AC\ {\isasymlongrightarrow}\ N\ {\isasymTurnstile}\ ZFC{\isacharparenright}{\isachardoublequoteclose}
\end{isabelle}
Here, \isatt{\isasymapprox} stands for equipotency, \isatt{nat} is the
set of natural numbers, and the predicate 
\isatt{Transset} indicates transitivity; and as usual, \isatt{AC}
denotes the Axiom of Choice, and \isatt{ZF} and \isatt{ZFC} the
corresponding subsets of \isatt{formula}.

%%% Local Variables: 
%%% mode: latex
%%% TeX-master: "forcing_in_isabelle_zf"
%%% ispell-local-dictionary: "american"
%%% End: 

\section{Related work}
\label{sec:related-work}

%% \textbf{Reviewer's comments}
%% {\it
%%   \begin{itemize}
%%   \item There, it would be appropriate to contrast what was done in
%%     Paulson's work on constructibility with the current work on forcing.
%%   \item More to the point, the recent work by Han and van Doorn on
%%     forcing in Lean deserves more discussion.  They have gone further
%%     than the current authors, having proved the independence of the
%%     continuum hypothesis.  They prefer Boolean-valued models as being
%%     more direct in use than the authors' countable transitive models.
%%     \begin{itemize}
%%     \item Readers will want to know whether the type-theoretic approach
%%       is better/worse/just different than using Isabelle/ZF, and
%%     \item are there any benefits to the ctm approach?
%%     \item Is the type-theory encoding of ZF really accurate?
%%     \item How about comparing proofs of equivalent statements in the two
%%       approaches for length and readability?
%%     \end{itemize}
%%   \end{itemize}
%% }

There are various formalizations of Zermelo-Fraenkel set theory in
proof assistants (v.g.\ Mizar, Metamath, and recently Lean
\cite{DBLP:conf/cade/MouraKADR15}) that proceed to different levels of
sophistication. Isabelle/ZF can be regarded as a notational variant of
NGB set theory \cite[Sect.~II.10]{kunen2011set}, because the schemes
of Replacement and Separation feature higher order (free) variables
playing the role of formula variables. It cannot be proved that the
axioms thus written correspond to first order sentences. For this
reason, our relativized versions only apply to set models, where
we restrict those variables to predicates that actually come
from first order formulas. In that sense, the axioms of the locale
\isatt{M{\isacharunderscore}ZF} correspond more faithfully to the
$\ZF$ axioms.

Traditional expositions of the method of forcing
\cite{kunen2011set,Jech_Millennium} are preceded by a study of
relativization and absoluteness. For this reason, it was a natural
choice at the beginning of this project to build on top of Paulson's
formalization of constructibility on Isabelle/ZF, and that was one
of the main early reasons to work on that logic instead of, e.g., HOL
---below we discuss other reasons. In any case, our
development of forcing does not depend on constructibility
itself (in contrast to Cohen's
original presentation, in which ground models are initial segments of the
constructible hierarchy).

A natural question is whether Isabelle/HOL (with a far more solid
framework to work with given its infrastructure and automation) would
have been a better choice than Isabelle/ZF. In fact,
there are two developments of Zermelo-Fraenkel set theory available on
it: \isatt{HOLZF} by Obua \cite{DBLP:conf/ictac/Obua06} and
\isatt{ZFC{\isacharunderscore}in{\isacharunderscore}HOL} by Paulson
\cite{ZFC_in_HOL-AFP}. But these (logically equivalent) frameworks are
higher in consistency strength than Isabelle/ZF. To elaborate on this,
both ZF and HOL are axiomatized on top of Isabelle's metalogic
\emph{Pure}, which is a version of ``intuitionistic higher order
logic.'' In  \cite{Paulson1989} Paulson proves that \emph{Pure}
is sound for intuitionistic first order logic, thus it does not add
any strength to it. On top of this, the axiomatization of Isabelle/ZF
results in a system equiconsistent with $\ZFC$. On the other hand,
showing the consistency of \isatt{HOLZF} (and thus
\isatt{ZFC{\isacharunderscore}in{\isacharunderscore}HOL}) requires
assuming the consistency of $\ZFC$ plus the existence of an
inaccessible cardinal \cite[Sect.~3]{DBLP:conf/ictac/Obua06}. We note,
in contrast, that our extra running assumption of the existence of a
countable transitive model is considerably weaker (directly and
consistency-wise) than the existence of an inaccessible cardinal.

Concerning the formalization of the method of forcing, to the best of
our knowledge there is only one other that deals with forcing for
set theory: the recent \emph{Flypitch} project by Han and van Doorn
\cite{han_et_al:LIPIcs:2019:11074,DBLP:conf/cpp/HanD20}, which
includes a formalization of the independence of CH using the Lean proof
assistant. The Flypitch formalization is largely orthogonal to ours
(it is based on Boolean-valued models, which are interpreted into
type theory through a variant of the Aczel encoding of set theory),
and this precludes a direct comparison of code. But we can highlight
some conceptual differences between our development and the
corresponding fraction of Flypitch.

A first observation concerns consistency strength. The consistency of
Lean requires infinitely many inaccessibles. More precisely, let
Lean$_n$ be the theory of CiC foundations of Lean restricted to $n$
type universes.  Carneiro proved in his MSc thesis~\cite{carneiro-ms-thesis} the consistency of Lean$_n$ from $\ZFC$ plus
the existence of $n$ inaccessible cardinals. It is also reported in
Carneiro's thesis that Werner's results in
\cite{10.5555/645869.668660} can be adapted to show that Lean$_{n+2}$
proves the consistency of the latter theory.  In that sense, although
Flypitch includes proofs of unprovability results in first order
logic, the meta-theoretic machinery used to obtain them is far heavier
than the one we use to operate model-theoretically.

In second place, a formalization of forcing with general partial
orders, generic filters and  ctms has ---in our opinion--- the added value
that this approach is used in an important (perhaps the greatest)
fraction of the literature, both in exposition and in research
articles and monographs. In verifying a piece of mature mathematics as the
present one, representing the actual practice seems paramount to us.
 
Finally, as a matter of taste, one of the main benefits of using transitive
models is that many fundamental notions are absolute and thus most of
the concepts and statements can be interpreted transparently, as we
have noted before. It
also provides a very concrete way to understand generic objects: as
sets that (in the non trivial case) are provably not in the original
model; this dispels any mystical feel around this concept (contrary
to the case when the ground model is the universe of all sets). In
addition, two-valued semantics is closer to our intuition.

%%% Local Variables: 
%%% mode: latex
%%% TeX-master: "forcing_in_isabelle_zf"
%%% ispell-local-dictionary: "american"
%%% End: 

\section{Conclusion and future work}
\label{sec:conclusion}

We consider that the formalization of the definition of $\forceisa$
and its recursive characterization of forcing for atomic formulas is
a turning point
in our project; the reason for this is that all further
developments will not involve such a daunting metamathematical
component. Even the proofs of the Fundamental Theorems of Forcing
turned out to follow rather smoothly after this initial setup was
ready, the only complicated affair being to show that various dense sets belong
to $M$. 
Actually, this is a point to be taken care of: For every new
concept that is introduced, some lemmas concerning 
relativization and closure must be proved to be able to synthesize its
internal definition. Further automation must be developed for this
purpose.

In the course of obtaining internal formulas for the atomic case of
forcing, a fruitful discussion
concerning complementary perspectives on the role of proof assistants
took place. An earlier approach relied more heavily in formula
synthesis, thus making the Simplifier an indispensable main
character. Following this line was quickier from the coding point of
view since few new primitives were introduced and thus fewer lemmas
concerning absoluteness and arities. On the downside, processing was a
bit slower, the formulas synthesized were gigantic and the process on
a whole was more error-prone. In fact, this approach was unsuccessful
and we opted for a more detailed engineering, defining all
intermediate steps. So the load on the assistant, in this part of the
development, balanced from code-production to code-verification. 

The next task in our path is pretty clear: To develop the forcing
notions to obtain the independence of $\CH$ 
along with the prerequisite combinatorial results, v.g.\ the
$\Delta$-system lemma. A development of cofinality is under way in a
joint work with E.~Pacheco Rodríguez, which is needed for a general
statement of the latter. Once these developments are finished, we will
be able to give a more thorough comparison between our project and
the \emph{Flypitch} approach using Boolean-valued models.

In a second release of \isatt{ZF-Constructible-Trans}, we intend to
conform it to the lines of \emph{Basic Set Theory (BST)} proposed by
Kunen~\cite[I.3.1]{kunen2011set} in which elementary results have
proofs using alternatively Powerset or Replacement. The interest in
this arises because many natural set models (rank-initial segments of
the universe or the family $H(\kappa)$ of sets of cardinality less
than $\kappa$ hereditarily) satisfy one of those axioms and not the
other. There are also still some older or less significant proofs
written in tactical (\textbf{apply}) format; we hope we will find the
time to translate them to Isar. Finally, the automation of formula
synthesis is on an early stage of development; finishing that module
will make writing our proofs of closure under various operations
faster, and also turn the set theory libraries more usable to other
researchers.

%%% Local Variables: 
%%% mode: latex
%%% TeX-master: "forcing_in_isabelle_zf"
%%% ispell-local-dictionary: "american"
%%% End: 


\providecommand{\noopsort}[1]{}

\end{document}